# Single-dislocation phonons: atomic-scale measurement and their thermal properties

Yue-Hui Li[1,2‡](李跃辉) Bo Han[1,2‡](韩博), Xiao-Long Yang[3‡](杨小龙), Rui-Lin Mao[1,2‡]*(毛瑞麟), Fa-Chen Liu[1,2](刘法辰), Ruo-Chen Shi[1,2](时若晨), Rui-Shi Qi[2](亓瑞时), Xiao-Rui Hao[2](郝小锐), Ning Li[2](李宁), Bing-Yao Liu[2](刘秉尧), Xiao-Mei Li[4](李晓梅), Jin-Long Du[2]*(杜进隆), Ji Chen[5,6,7](陈基), Wu Li[8]*(李武), Peng Gao[1,2,6,7]*(高鹏)

[1]*International Center for Quantum Materials, School of Physics, Peking University, Beijing, 100871, China*

[2]*Electron Microscopy Laboratory, School of Physics, Peking University, Beijing, 100871, China*

[3]*College of Physics, and Center of Quantum Materials and Devices, Chongqing University, Chongqing 401331, China*

[4]*School of Integrated Circuits East China Normal University Shanghai 200241, China*

[5]*Institute of Condensed Matter and Material Physics, School of Physics, Peking University, Beijing, 100871, China*

[6]*Interdisciplinary Institute of Light-Element Quantum Materials and Research Center for Light-Element Advanced Materials, Peking University, Beijing 100871, China.*

[7]*Collaborative Innovation Centre of Quantum Matter, Beijing 100871, China*

[8]*Eastern Institute for Advanced Study, Eastern Institute of Technology, Ningbo 315200, China*

Yue-Hui Li, Bo Han, Xiao-Long Yang and Rui-Lin Mao contributed equally to this work.

[*]Corresponding author. E-mail: ruilinmao@stu.pku.edu.cn; jldu@pku.edu.cn; wu.li.phys2011@gmail.com; pgao@pku.edu.cn

**Abstract** Nanoscale defects such as dislocations, have a significant impact on the phonon thermal transport properties in non-metallic materials. To unravel these effects, understanding of defect phonon modes is essential. Herein, at the atomic scale, the localized phonons of individual dislocation at a Si/Ge interface are measured via monochromated electron energy loss spectroscopy in a scanning transmission electron microscope. These modes are then correlated with the local microstructure, further revealing the dislocation effects on the local thermal transport properties. The dislocation causes phonon redshift in several milli-electron-volts within about two to four nanometers of the core, where both of the strain field and Ge-segregation play roles. With the presence of dislocation, the local interfacial thermal conductance can be either enhanced or reduced, depending on the complex interaction and competition between lattice-disorder (dislocation) and element-disorder (heterointerface mixing and Ge-segregation) at the interface. These findings provide valuable insights to improve the thermal properties of thermoelectric generators and thermal management systems through proper defect engineering.

The atomic-scale and nanoscale lattice imperfections such as heterointerfaces, grain boundaries and dislocations, usually have unique phonon scattering behaviors different than the bulk matrix and thus a significant impact on thermal properties[1,2] in the non-metallic systems such as thermal management materials[3,4] and thermoelectric devices[5,6]. Therefore, probing the local phonon modes and scattering behavior at these imperfections/interfaces is the key to understand the underlying physics of their thermal properties (e.g., interfacial thermal conductance (ITC)[7]) and subsequently gain insights into the optimization. Indeed, numerous strategies have been reported to tune the ITC by introducing intermediate matching layer[8], ion irradiation[9,10], dislocations[11,12] and so on. For example, the dislocation is generally considered to increase local thermal resistance. The underlying mechanism lies in that local lattice disorder at dislocations induced by translation symmetry breaking — which refers to deviations from an ideal crystalline arrangement, encompassing point defects (e.g., vacancies), line defects (e.g., dislocations) and planar defects (e.g., stacking faults) — can strongly scatter the bulk phonons, thereby reducing the phonon lifetime. Previous studies reported that the dislocations can almost reduce the ITC by half across the GaSb/GaAs interface[11]. Dislocations within grain boundaries of PbSe significantly enhance the phonon scattering, thereby leasing to a remarkably improved thermoelectric performance[13]. The lattice strain fluctuations induced by in-grain dislocation notably broaden the phonon dispersion and substantially reduce the ITC, thus enabling an extraordinary figure of merit of ~2.5 in PbTe alloys[14]. However, in addition to lattice disorder, element disorder also usually exists at the dislocation even at sub-nanometer scale[15]. Element disorder denotes a non-ideal state of random or uneven atomic distribution in materials, manifesting as random lattice substitution (e.g., irregular atomic site occupancy in solid solutions), segregation (local atomic clustering), vacancies, or compositional fluctuations—ultimately leading to loss of periodicity in nanometer-to-atomic-scale composition. Such element-disorder should also play an important role in the interfacial thermal transport, e.g., a higher ITC across Al/Si interface can be achieved after ion irradiation due to the ion mixing of the native oxide and silicon[10]. Therefore, the defects in practical materials with multiple disorders can influence the local thermal transport properties in a very complicated manner due to the competitions among multiple factors. Revealing such information is highly desirable.

However, most of previous works on ITC are based on either theoretical calculations or macroscale thermal transport experiments. To accurately understand effects of defects on phonon transport properties, the information of local phonon modes associated with the specific lattice structure and atomic arrangement of individual defects is required. In this work, we combine scanning transmission electron microscopy (STEM), electron energy loss spectroscopy (EELS) and molecular dynamics (MD) simulations to investigate the localized phonon modes of individual dislocations at a Silicon-Germanium (Si/Ge) interface. The recent progress in EELS equipped with a monochromator in an aberration corrected STEM has made it possible to measure localized vibrational spectra at nanoscale and atomic-scale[16-26] and even correlate the localized defect phonons with thermal transport properties[27-29]. For example, recent works reported that the grain boundaries of silicon[30], $SrTiO_3$[31] , and $Al_2O_3$[32] significantly change local phonon density of states (PhDOS). Meanwhile, the distribution of strain and elements can be simultaneously obtained through quantitatively determining the local atomic arrangement under the STEM studies. The chosen system, Si/Ge alloy, is an excellent thermoelectric material with wide applications in thermoelectric generators converting heat flows into usable electrical energy[33,34] and thermoelectric microrefrigerators providing large cooling power densities[35], e.g., the thermal energy of radioactive heat sources[36]. Several methods have been proposed to engineer the thermoelectric properties of SiGe alloys, such as nanostructured SiGe alloys[37], SiGe/Si superlattices[38], defects[39], etc. However, the localized phonons of the dislocation in Si-Ge alloy system and their effects on the thermal transport properties are still largely unknown.

We find that the interface dislocation strongly alters the local phonon modes and thermal conductance. Within about four nanometers of the dislocation core, localized phonon modes exhibit a redshift of several milli-electron-volts (lower than bulk Si and Ge phonon energies), stemming from combined strain fields and elemental segregation. Specifically, tensile strain at the Ge side drives phonon red-shift, whereas Ge segregation at the dislocation core dominates phonon softening at the Si side. Although it's widely recognized that the lattice disorder in single crystals decreases the thermal conductance, the effects of dislocation at the Si/Ge interface on thermal transport is surprisingly complex—arising from the interplay and competition between the lattice-disorder (discontinuous lattice structure) and element-disorder (heterointerface mixing and Ge-segregation) mechanisms. Our findings indicate that depending on the elemental segregation, the local ITC around the dislocation can be either enhanced or reduced. For an atomically sharp heterointerface, where substantial phonon energy mismatches exist, the lattice disorder induced by dislocations effectively bridges the PhDOS between the two materials, enhancing PhDOS overlap. Conversely, as the segregation and interdiffusion increase, the system evolves toward a quasi-homogeneous, single-crystal-like state: phonon energy disparities at the interface diminish, reducing heterointerface's contribution to thermal resistance, while the role of dislocation-associated lattice-disorder becomes dominant. In our case, pronounced Ge-segregation in the dislocation core leads to ITC degradation akin to single crystal systems, where dislocation-induced phonon scattering becomes the primary resistance mechanism. Our study unveils the localized phonon modes of individual dislocations within complex mechanical-chemical environments, offering critical insights for designing advanced devices with tailored thermal properties via proper dislocation engineering.

*Atomic structure and EEL spectra near the dislocation.* Figure 1(a) shows an atomically resolved high-angle annular dark-field (HAADF) image of a Si/Ge heterojunction interface. The lattice constant of Ge, $d_{Ge}$=5.66 Å, is approximately 4% larger than that of Si, $d_{Si}$=5.43 Å, leading to formation of misfit edge dislocations at the interface as shown in Fig. 1(b). Based on the element distribution maps from core-loss EELS measurement in the Fig. 1(c), Fig. S1 and the Z-contrast in Fig. 1(b), the element interdiffusion between Ge and Si occurs to form a disordered SiGe mixing layer with a thickness of about two nanometers at the dislocation-free interface, while significant Ge segregation takes place at the dislocation core region.

Figure 1(d) shows the acquired EEL spectra in bulk Si, bulk Ge, interfacial mixing layer at a dislocation-free region, and interfacial dislocation core by using off-axis geometry measurement[40], which is beneficial for enhancing the signal to zero loss peak (ZLP) ratio especially for the low energy phonons (see Fig. S2). The spectra are averaged from a spatially resolved map with the standard errors. The details of data processing are shown in Methods and Fig. S3. In order to identify the spectral features and correlate them with vibrational modes, the projected phonon density of states (PPDOS) and phonon dispersion were calculated and shown in Fig. 1(e) and Fig. S4 based on density functional perturbation theory (DFPT) (see details in Methods). The calculated phonon modes are in good agreement with those in previous study[41]. To simplify the description of experimental data, the peaks of phonon spectra are classified into three branches, labeled as $\alpha$ (arrows without a tail), $\beta$ (arrows with a tail) and $\gamma$. According to the calculated PhDOS and phonon dispersion of Si and Ge, $\alpha$, $\beta$ and $\gamma$ are mainly assigned to transverse acoustic (Si-TA) mode, longitudinal acoustic/optical (Si-LAO) mode and transverse optical (Si-LO/TO) mode in Si side, respectively. In Ge layer, $\alpha$ and $\beta$ are assigned to Ge-TA mode and Ge-AO (LA, LO, and TO all contribute) mode, respectively. Since the Si-LO/TO phonon is pronounced only in the Si layer, $\gamma$ is absent in the Ge layer. At the dislocation-free interface region, the $\alpha$ and $\beta$ peaks are between those in bulk Si and bulk Ge due to element interdiffusion and/or heterointerface phonon modes[23], while at the dislocation core they have lower energy than either of them (i.e. redshift), indicating the generation of new localized defect modes with different phonon energy.

*Localization and redshift feature of the dislocation phonons.* Figure 2(a) shows energy-filtered EELS intensity maps of window 6-14 meV, in which the phonon spectrum should be mainly attributed to the dislocation based on Fig. 1(d). It can be observed that the integrated intensity is concentrated at the dislocation core, further verifying the redshift of $\alpha$-peak at the dislocation. To quantitatively validate these spectral features, the background-subtracted EEL spectra were fitted using a simple Gaussian peak fitting model. The data processing details are included in the Methods. The energy distribution maps of $\alpha$, $\beta$ and $\gamma$ phonon branches are shown in Figs. 2(b-d), respectively. Notably, near the dislocation core, three phonon branches all show redshift. In contrast, at the dislocation-free interfacial region, the phonon energies of $\alpha$ and $\beta$ modes gradually transform from Si on the left to Ge on the right, which act as phonon bridges to connect significantly different energies of phonons[29,42]. Specifically, Figure 2(e) shows the extracted line profiles of $\alpha$, $\beta$ and $\gamma$ phonon energy through the interface dislocation (red) and dislocation-free region (black) perpendicular to the interface denoted by the solid rectangles in Fig. 2(a), where all of the dislocation-induced localized modes, $\alpha_{\mathrm{disl}}$, $\beta_{\mathrm{disl}}$, and $\gamma_{\mathrm{disl}}$ modes show redshift at the dislocation core (labeled as 0 nm). The energy of $\alpha_{\mathrm{disl}}$ mode is ~1.2 meV lower than that of bulk Ge-TA mode, the energy of $\beta_{\mathrm{disl}}$ mode is ~3.1 meV lower than that of bulk Ge-AO mode, and the energy of $\gamma_{\mathrm{disl}}$ mode is ~2.3 meV lower than that of the regular interface mode. Figure 2(f) show the phonon energy changes of $\alpha$, $\beta$ and $\gamma$ modes through the dislocation core along the interface of the full acquisition area denoted by the rectangles in Fig. S5. Therefore, the interfacial dislocation has very different localized phonon modes than the dislocation-free region, implying its important influence on the thermal transport properties. The corresponding Gaussian fitted line profiles have a full width of half maximum of ~2.5 nm for $\alpha_{\mathrm{disl}}$, ~3.2 nm for $\beta_{\mathrm{disl}}$ and ~4.4 nm for $\gamma_{\mathrm{disl}}$ respectively, indicating highly localized nature of the dislocation phonon modes. In fact, due to the possible collection of dipole signals, the Umklapp processes[43], and the de-channeling effects[44], the estimated delocalization range of dislocation signals by using large convergence may be slightly exaggerated, i.e., the true physical ranges of dislocation phonons are even smaller. Moreover, the width of low energy phonon $\alpha$ is the smallest likely due to the impact nature of acoustic phonons compared to the optical ones[23]. To verify the behavior of redshift and localization characteristic of dislocation phonons, frequency resolved frozen phonon multislice (FRFPMS)[45] simulations were also performed. Figure S6(a) showed the calculated scattering cross-section with the same experiment conditions (see details in the Methods) while the energy filtered cross-section mappings were shown in Figs. S6(b-d). The redshift of three peaks is well distinguishable and the dislocation phonons are highly localized, which in good agreement with the experimental data. Note that the phonon activity has been considered in this simulation, i.e., these local dislocation phonon modes in the simulation must be excitable by the electrons in this case.

*MD simulations and thermal transport properties of the Si-Ge interface.* To further understand the origin of dislocation phonons and their effects on thermal transport properties, MD simulations are performed. Figure 3(a) shows the calculated PPDOS of Ge single crystal without (black) and with (red) an edge dislocation, respectively. Such an edge dislocation produces complex strain fields nearby as illustrated by the experimental strain maps (Fig. S7). Based on simulations in Fig. 3(a), the tensile strain region (the third layer) shows a red shift of $\alpha_{\mathrm{disl}}$, $\beta_{\mathrm{disl}}$ (as well as the redshift of $\gamma_{\mathrm{disl}}$ modes in Fig. S8), while the compressive strain region (the second layer) causes a blue shift of $\beta_{\mathrm{disl}}$ and $\gamma_{\mathrm{disl}}$ instead. However, only red-shift is observed in our experiment. This discrepancy between calculation and experiment suggests that the strain condition is not the sole mechanism. Indeed, the significant element segregation at the dislocation core (as shown in Fig. 1(c)) should also play an important role. At the Si side with compressive strain, the phonon energy is expected to blue-shift, but the significant Ge segregation (as shown in Fig. 1(c)) at the dislocation core tend to make the phonon red-shift as the presence of heavier Ge atoms can lower the phonon energies. In this sense, the phonons at the dislocation rely on the competition between the lattice-disorder and element-disorder. Apparently, the Ge segregation is dominated at the Si side. This is also validated by the experimental decoupling of the element segregation and strain contribution to the energy shift, the details are illustrated in Fig. S9.

Now we discuss how the thermal transport properties are influenced by the dislocation and interface disorder layer with the aid of simulations. In general, the dislocations in a single crystal can cause a reduction in the phonon group velocity[46] and an increase in the anharmonic phonon-phonon scattering[47]. Both effects lead to the reduction of ITC[12], which is beneficial for the enhancement of thermoelectric performance. For example, previous study reported that the scattering of a wide spectrum of phonons by dislocations, grain boundaries and other imperfections in SiGe alloys leads to a significantly low thermal conductivity ~0.93 W/(m·K), and thus enhance the thermoelectric figure-of-merit ~1.84 at 1073 K[48]. In fact, besides SiGe alloys, dislocations were also introduced into many other thermoelectric materials to achieve high performance[49,50].

However, the case for the dislocation at an heterointerface as studied here is more complicated. It should be noted that the mixing layer plays an important role in the phonon transport across an heterointerface. Previous study reported the existence of localized interfacial phonon modes at the Si/Ge interface having a significant contribution to the interfacial thermal conductance[27]. Since there exists significant phonon mismatch, many phonons (particularly those of optical phonons in Si) necessitate energy transfer to the other side through inelastic scattering processes in which the localized interfacial phonon modes play a facilitating role. The inelastic scattering redistributes thermal energy among phonons through non-equilibrium effects induced by the heterojunction[51,52]. With the mixing layer, our study indicates that the low-frequency phonon modes surrounding $\alpha$ and $\beta$ peaks act as bridges to connect mismatch in phonon energy at the interface[29,42,53], as shown in Fig. 2(e), which is beneficial to increase the ITC. In our practical case with both of dislocation and interdiffusion at the interface, the effects of dislocation on thermal transport contains not only their competitions but also their interactions. To quantify the effect of these factors, we calculate the room-temperature lattice thermal conductivity ($\kappa$) of different structures with and without inclusion of dislocation and interdiffusion, whose corresponding spectral thermal conductivities are shown in Figs. 3(c-f). For comparison, the case of the pure Ge single crystal is also given. From Fig. 3(c), the dislocation in pure Ge single crystal reduces its $\kappa$, as broadly seen in other single crystal systems[47,54]. For the Si/Ge interface without mixing layer, the dislocation leads to $\kappa$ increasing instead. As for the interface with intermixing, the change of thermal conductivity is quite complicated, highly depending on the elemental distribution at the interface (the corresponding atomic models are shown in Fig. 3(b) and Fig. S10). The presence of dislocation increases $\kappa$ in the case with the most significant interdiffusion (50% element mixing) but reduces $\kappa$ in the case with Ge-segregation in the dislocation core in our case.

The interfacial thermal transport behaviors above can be understood by examining the local DOS near the interface. Generally, the number of heat channels across an interface is closely related with the PhDOS overlap between the two sides of the interface according to the diffuse mismatch model, i.e., the higher (lower) the PhDOS overlap, the more (less) phonon transport channels[53]. The dislocation in pure Ge single crystal causes the vibrational mismatch of its surrounding atoms, thereby suppressing its κ in a wide frequency range (Fig. 3(d)). For the Si/Ge interface without mixing layer, the dislocation induces the redshift of phonons near the interface, which can weaken the vibrational mismatch caused by the heterointerface, i.e., increase the PhDOS overlap, as listed in Table S1, thus giving rise to the increased contribution of mid- and low-frequency phonons (0-27 meV) (Fig. 3(e)). When elemental interdiffusion is introduced into the interface, the element disorder bridges the phonon DOS between two sides of the heterointerface and thus increases the PhDOS overlap. For the case of significant Ge-segregation in particular, the interface effects are severely diminished by the element disorder. In this case, the effect of introducing dislocations is similar to the single crystal case, which enables the suppression of κ in a wide frequency range as confirmed in Fig. 3(f). Figure 3(g) qualitatively illustrates the competitive relationship between dislocations and elemental mixing, where yellow (purple) hues indicate lower (higher) ITC. For Si/Ge interface without mixing layer, where there is a significant phonon energy mismatch between the two sides, dislocations can enhance inelastic phonon scattering, thereby improving ITC. However, when elemental mixing is introduced at the interface, the phonon energy difference between the two sides decreases, weakening the promoting effect of dislocations on inelastic phonon scattering and reducing the extent of ITC enhancement. When the interface contains a high concentration of Ge, the system more closely resembles a single crystal, with minimal phonon energy difference across the dislocation. In this case, the elastic scattering of phonons by dislocations becomes dominant, leading to a reduction in ITC.

In summary, we directly measure the localized phonon modes of dislocations at the Si/Ge interface using atomic resolution STEM-EELS and reveal the effects on local thermal conductivity with the aid of MD simulations. Dislocations induce highly localized phonon modes within two nanometers of the core, exhibiting a redshift of several milli-electron-volts, which are lower than those in bulk Si and Ge. The competition between strain and element segregation account for the energy shift. Notably, the dislocation influences the thermal transport in a complex manner, i.e., depending on the degree of element-disorder the local ITC could be either enhanced or reduced with the presence of dislocations, indicating intricate interactions between the lattice-disorder (structural discontinuities) and element-disorder (compositional heterogeneity). These findings provide useful information to modulate thermal properties via defect engineering.

*Materials And Methods*

*TEM sample preparation.* The cross-sectional TEM specimen was prepared by mechanical polishing followed by an argon ion milling carried out using the Precision Ion Polishing System (Model 691, Gatan Inc.). The specimen was baked at 160 ºC for 16 hours to further remove the surface amorphous layer before EELS experiments.

*Experimental setup.* The vibrational spectra were acquired at a Nion U-HERMES200 electron microscope equipped with both the monochromator and the aberration corrector operated at 60 kV. The convergence semi-angle and the collection semi-angle were both 25 mrad. The electron beam was moved off optical axis with ~50 mrad for off-axis experiments[40]. The scattering geometry and diffraction disks are shown in Fig. S11. The energy dispersion channel was set as 0.5 meV with 2048 channels in total.

To exclude the possible effects from diffraction effects, slit aberration and aperture charging etc. on the signal below 20 meV, a set of 2D-EELS images were collected near the interface (Fig. S12). The 2D-EELS data was collected under the same experimental conditions with Fig. 1(d), i.e., the convergence semi-angle and the collection semi-angle were 25 mrad. The beam was shifted ~50 mrad (off-axis) along the interface.

*DFPT simulations.* The PhDOS were calculated using density functional perturbation theory by Quantum ESPRESSO[55]. The ultra-soft pseudopotential (USPP)[56] and the Perdew-Burke-Enzerhof for solids (PBEsol)[57] exchange-correlation functional were used. The kinetic energy cut-off was 50 Rydbergs (Ry) for wavefunctions and 500 Ry for charge density and potential. The phonon dispersion and PhDOS was calculated by interpolating the dynamical matrix on an $8 \times 8 \times 8$ q-mesh.

*MD simulations.* The MD simulations were performed using the Large-scale Atomic/Molecular Massively Parallel Simulator (LAMMPS) package[58] with a velocity Verlet algorithm for numerical integration of the equations of motion and a time step of 1 fs. All the interatomic interactions were described by the Stillinger-Weber potential [59,60]. The simulation systems contain 100 × 10 × 2 conventional cells on both Si and Ge sides. Periodic boundary conditions were applied in all directions. To create the random atomic mixing layer, we selected the 4-unit-cell-long Ge layer near the interface and randomly replace several Ge atoms as Si atoms. We generated a series of interface configurations with random atomic mixing layer and obtained the most energetically stable atomic configuration by simulating the anneal-to-quench process from 1000 to 300 K. Our simulations were carried out in three steps. First, we performed simulations at 300 K and zero pressure in the isothermal-isobaric (NPT) ensemble using the Nosé/Hoover thermostat for 50 ps to help the system reach thermal equilibrium. Then the simulations were switched to the micro-canonical (NVE) ensemble for a further 50 ps to calculate the velocity autocorrelation function (VACF). Finally, nonequilibrium molecular dynamics (NEMD) simulations were conducted to calculate thermal conductivities of six simulation systems for another 200 ps in the NVE ensemble, which is long enough to bring the systems to the stationary state and obtain the corresponding heat flux ($J$) and temperature gradient along the $x$ direction ( $\nabla T$). A fixed boundary condition is applied along the heat flow direction to eliminate direct thermal coupling between the heat source and sink regions. The temperatures at the source and sink regions of the simulation domains were controlled with Langevin thermostats [61,62]. Figure S13 show the temperature profiles of NEMD simulations and accumulated heat of the heat source domain and heat sink domain varying as a function of the simulation time.

The PhDOS was calculated by the fast Fourier transformation of VACF. The PhDOS projected on atom *j* can be calculated:

$$g_j(\omega) = \int e^{i\omega t} \frac{\langle v_j(t) v_j(0) \rangle}{\langle v(0) v(0) \rangle} dt$$

where ω is the phonon frequency, $v_j$ is the velocity of atom *j*, the $\langle \cdots \rangle$ brackets indicate averaging over different starting step along the trajectory. The entire simulation domain was divided into many cells, and the phonon density of states contributed by the atoms of each cell were calculated. According to the Fourier's law, the lattice thermal conductivity (κ) is expressed as:

$$\kappa = -\frac{J}{\nabla T}.$$

To quantify the effects of dislocation and random atomic mixing on the thermal transport of Si/Ge systems, we also calculated the spectral thermal conductivity as:

$$\kappa_{sp}(\omega) = -\frac{q(\omega)}{A \nabla T},$$

where $q(\omega)$ is the frequency-dependent spectral heat current, and *A* is the cross-sectional area of the simulation domain. More computational details and full expression for $q(\omega)$ are referred to Refs [52,62,63]. Note that herein the temperature gradient was directly calculated as $\Delta T/L$ with a temperature difference between thermal reservoirs of $\Delta T$ and the simulation domain length of *L*, which has been demonstrated to be effective and reliable in computing κ[64].

*FRFPMS simulations.* Frequency resolved frozen phonon multislice method proposed by Zeiger et al.[45] was performed to verify whether the local modes near the dislocation can be excited by the electron beam. Model of 8 nm×16 nm×0.8 nm with the same dislocation structure used in thermal properties calculations was used to calculate local vibration spectrum. The system was fully relaxed under NPT ensemble and then in NVE ensemble. A 100 ps length NVE trajectory with sample interval of 20 fs was used to calculate band filtered trajectory. We performed Fast Fourier Transformation (FFT) on NVE MD trajectories in time domain then filtered trajectories with 1 THz width Gaussians window-based bandpass filter with different central vibrational frequency to achieve mode-specific refinement. This explicit parameterization (window function type, bandwidth magnitude) constitutes our unique spectral refinement strategy within the broader bandpass filtering framework. The frozen phonon multislice calculations were performed using MD Frozen Phonon module embedded in abTEM package[65] with slice thickness of 0.2 nm. The statistical factor[66] of phonon population under 300 K is multiplied to suppress the excessive strong phonon signal under 20 meV[67,68]. 10 meV width Gaussians were convoluted to smooth the spectra and 3 meV width Gaussians were used in spectrum image to guarantee the visibility of localized phonon modes.

*Phonon EEL spectra processing.* The phonon EEL spectra are firstly normalized by the integral of zero-loss peaks (ZLP) , aligned using cross correlation method then denoised using BM3D algorithm[69]. For background subtraction, the ZLP are fitted using Pearson VII function[70] with the energy window of [-6, 6] meV and [79, 99] meV. The function of Pearson VII is shown as Eq1:

$$I(2\theta) = I_{max} \frac{w^{2m}}{[w^2 + (2^{1/m} - 1)(2\theta - 2\theta_0)^2]^m}$$

Where $I_{max}$ is the peak value, w is the width of the peak, and m is the shape factor. The peak becomes a Lorentzian while m→1 and a Gaussian when m→0.

The typical full width at half maximum (FWHM) of the fitted ZLP is ~ 20 meV.

The ZLP subtracted data is then fitted using three Gaussian peaks with the fitting window of [10, 30] meV for $\alpha$-peak, [20, 50] meV for $\beta$-peak and [30, 70] meV for $\gamma$-peak. Figure S14(a) shows the multi-gaussian fitting results of the EEL spectra acquired at different positions and Figs. S14(b-d) show the stand deviation error, the Gaussian fitting error and their combined uncertainty for peaking fitting of phonon energies related to Figure 2. While we note that the Gaussian fitting error is non-negligible, the characteristic redshift of the γ peak remains clearly resolvable and statistically significant. This observation is robust against the fitting uncertainties, as confirmed by our error analysis.

*Core-loss EEL Spectra processing***.** Si and Ge elemental maps are extracted from Si-*K* edge and Ge-*L* edge of core-loss EELS. The background baselines of core-loss EEL spectra are subtracted using power law with the fitting window of [1430, 1860] eV for Si and [1120, 1200] eV for Ge. To extract the Si and Ge elemental maps, the energy windows for intensity integral on EEL spectra are [1871.0, 1890.6] eV for Si-*K* edge and [1116.4, 1402.1] eV for the Ge-*L* edge.

*GPA analysis.* The geometry phase analysis (GPA) was performed by FRWRtools in DigitalMicrograph software, based on the work done by Hytch et al.[71] The FRWRtools available from https://www.physics.hu-berlin.de/en/sem.

*Supplementary Materials.* The following files are available free of charge. Supplementary Materials.docx. Validation of EELS acquisition and data processing methods; Elemental and strain distribution near dislocation core and the decoupling of their effects on phonon energy shift; Additional MD simulations on Si dislocations, DOS overlap and detailed simulation models; FRFPMS simulation results.

*Author Contributions.* P.G. conceived and supervised the project. Y. H. L., B. H., R. C. S., N. L. and X. M. L performed the initial research under direction of J. L. D. and P.G.; X. L. Y. and R. L. M. performed simulations under direction of J. C. and W. L.; F. C. L., R. S. Q. and R. C. S. assisted the data analysis. X. R. H. and B. Y. L. fabricated the TEM sample. Y. H. L., B. H. and R. L. M wrote the primitive manuscript under the direction of P.G. with contributions from all other co-authors. ‡These authors contributed equally.

*Acknowledgements.* This work was supported by the National Natural Science Foundation of China (52125307) and the National Key R&D Program of China (2021YFB3501500). P.G. acknowledges the support from the New Cornerstone Science Foundation through the XPLORER PRIZE. We thank the helpful discussion with Prof. Jiaqing He at Southern University of Science and Technology (China) and Prof. Eric Stach at University of Pennsylvania (USA). We acknowledge the Electron Microscopy Laboratory of Peking University for the use of electron microscopes. We acknowledge the High-performance Computing Platform of Peking University for providing computational resources for simulations.

## Figures and captions

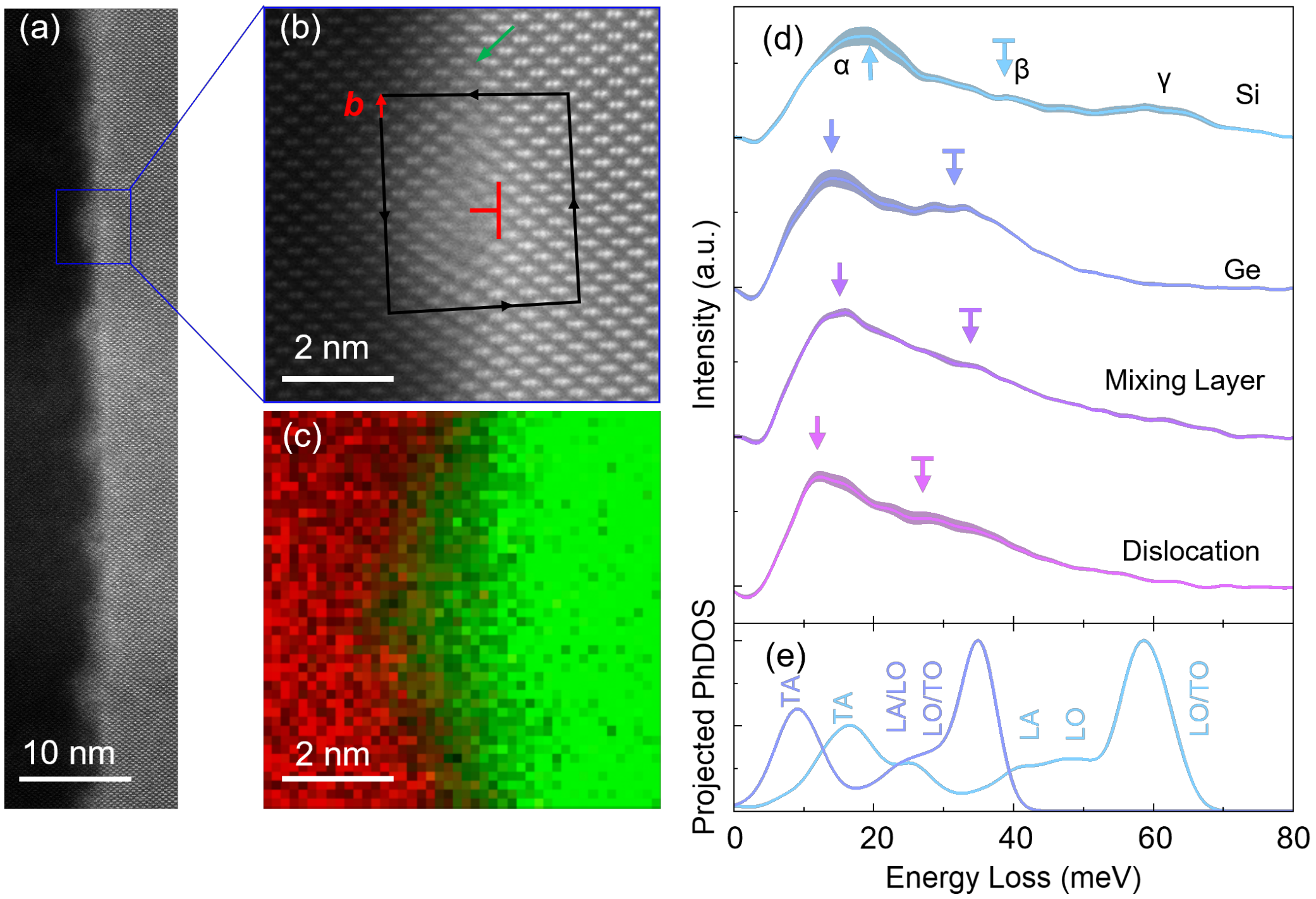

**Fig. 1 HAADF image and EEL spectra acquired at different locations.**

**(a)** A HAADF image of a Si/Ge interface. **(b)** An enlarged view showing the atomic structure of an interface dislocation. The Burgers vector ***b*** is denoted by the red arrow. The green arrow denotes the disordered mixing layer at the dislocation-free interface region based on the Z-contrast. **(c)** Element distribution of Si (red) and Ge (green) at the dislocation extracted from EELS of Si-*K* edge and Ge-*K* edge. **(d)** EEL spectra with standard error acquired at different locations. The arrows without (with) tail denote $\alpha$ ($\beta$) peaks. The peak assignment is conducted by combing spectroscopic features with theoretical simulation results. The error bar is defined by the standard deviations between binned 5×5 pixels on different places of the sample. The spectrum was normalized by its integral before ZLP subtraction. **(e)** Calculated phonon PhDOS projected on [110] plane of Si (bule) and Ge (purple).

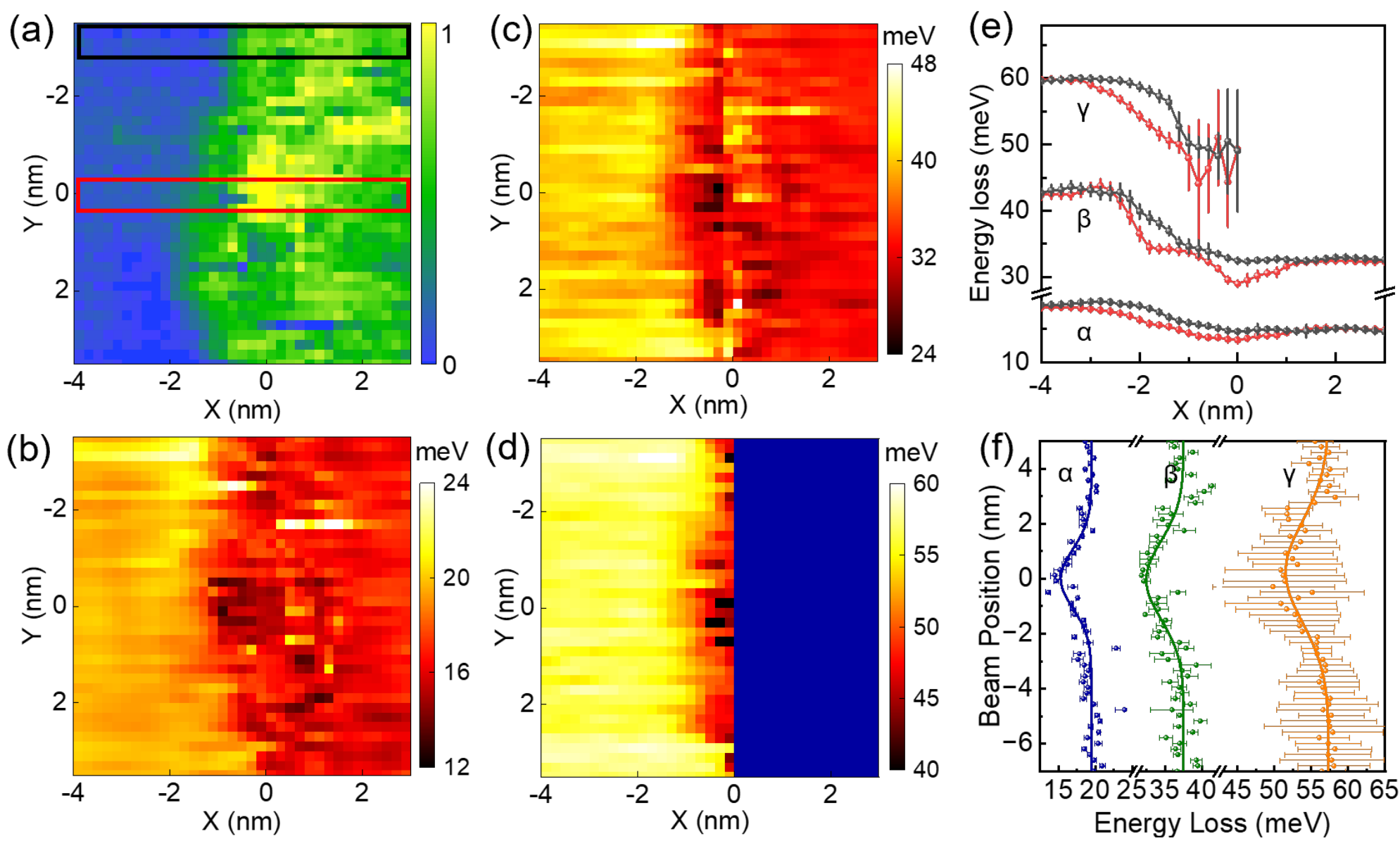

**Fig. 2 The intensity and energy distribution maps of phonons.**

**(a)** The energy-filtered EELS intensity map of 6-14 meV window corresponding to one branch of the dislocation phonon modes. **(b-d)** The energy distribution maps of $\alpha$, $\beta$ and $\gamma$ phonon branches. The phonon energy is extracted by a simple Gaussian peak fitting model. All of them showing the lower energy at the dislocation. **(e)** The line profiles of $\alpha$, $\beta$ and $\gamma$ phonon energy through the mixing layer (black) and interface dislocation (red) along the X direction denoted by solid rectangles in a. **(f)** The data points of the phonon energy of $\alpha$, $\beta$ and γ modes through the dislocation core along the Y direction denoted by rectangles in Fig. S5. The solid lines are the fitted Gaussian distributions. The error bar is defined by the sum of standard deviations error and Gaussian fitting error, see Methods for more information.

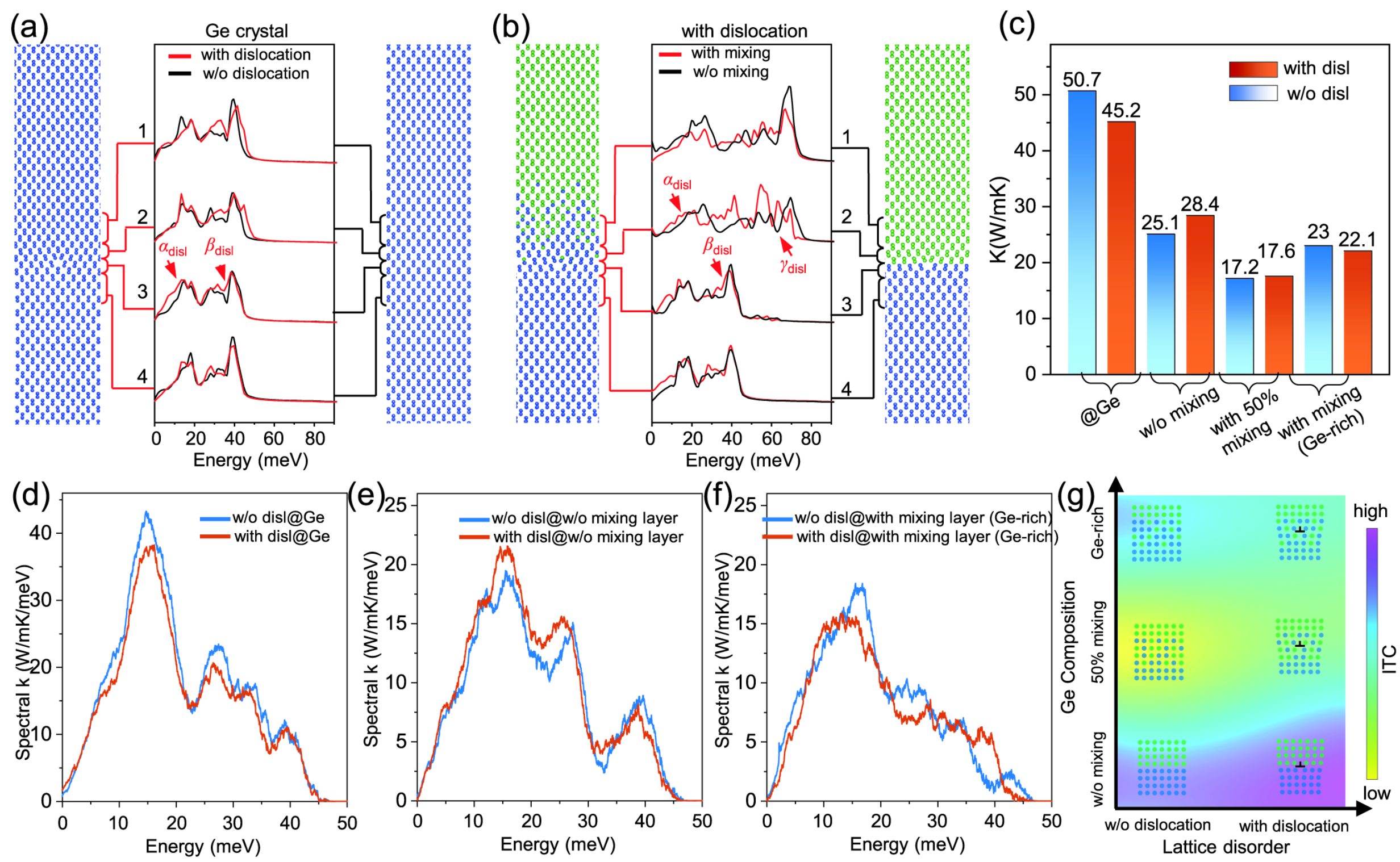

**Fig. 3 Molecular dynamics simulations of different Si/Ge systems.**

**(a)** Schematic diagrams and calculated PhDOS of Ge crystal. Each layer is the PhDOS of the atoms covered by the brace. The dislocation modes $\beta_{\text{disl}}$ (the red arrow) experience a redshift compared to dislocation-free case (black). **(b)** Schematic diagrams and calculated PhDOS of Si/Ge systems with a dislocation. The dislocation modes show redshift at the intermixing region. **(c)** The dislocation effect on thermal conductivity in Ge, Si/Ge without mixing layer, Si/Ge with a 50% intermixing layer, and Si/Ge with significant Ge segregation. Although in a single crystal Ge, the dislocation reduces the ITC, while at the Si/Ge interface, the dislocations (orange) can either enhance or reduce the ITC compared to the dislocation-free case (blue). **(d)** The spectral thermal conductivity of Ge in the case of with dislocation (orange) and without dislocation (blue). **(e)** The spectral thermal conductivity of Si/Ge system without mixing layer. **(f)** The spectral thermal conductivity of Si/Ge system with mixing layer (in a Ge rich case). **(g)** Schematic diagram of the competitive relationship between dislocations and elemental mixing.

Supporting Information for

**Single-dislocation phonons: atomic-scale measurement and their thermal properties**

Yue-Hui Li[1,2‡](李跃辉) Bo Han[1,2‡](韩博), Xiao-Long Yang[3‡](杨小龙), Rui-Lin Mao[1,2‡]*(毛瑞麟), Fa-Chen Liu[1,2](刘法辰), Ruo-Chen Shi[1,2](时若晨), Rui-Shi Qi[2](亓瑞时), Xiao-Rui Hao[2](郝小锐), Ning Li[2](李宁), Bing-Yao Liu[2](刘秉尧), Xiao-Mei Li[4](李晓梅), Jin-Long Du[2]*(杜进隆), Ji Chen[5,6,7](陈基), Wu Li[8]*(李武), Peng Gao[1,2,6,7]*(高鹏)

[1]*International Center for Quantum Materials, School of Physics, Peking University, Beijing, 100871, China*

[2]*Electron Microscopy Laboratory, School of Physics, Peking University, Beijing, 100871, China*

[3]*College of Physics, and Center of Quantum Materials and Devices, Chongqing University, Chongqing 401331, China*

[4]*School of Integrated Circuits East China Normal University Shanghai 200241, China*

[5]*Institute of Condensed Matter and Material Physics, School of Physics, Peking University, Beijing, 100871, China*

[6]*Interdisciplinary Institute of Light-Element Quantum Materials and Research Center for Light-Element Advanced Materials, Peking University, Beijing 100871, China.*

[7]*Collaborative Innovation Centre of Quantum Matter, Beijing 100871, China*

[8]*Eastern Institute for Advanced Study, Eastern Institute of Technology, Ningbo 315200, China*

Yue-Hui Li, Bo Han, Xiao-Long Yang and Rui-Lin Mao contributed equally to this work.

[*]Corresponding author. E-mail: ruilinmao@stu.pku.edu.cn; jldu@pku.edu.cn; wu.li.phys2011@gmail.com; pgao@pku.edu.cn

# S1 Supporting Information: Figures

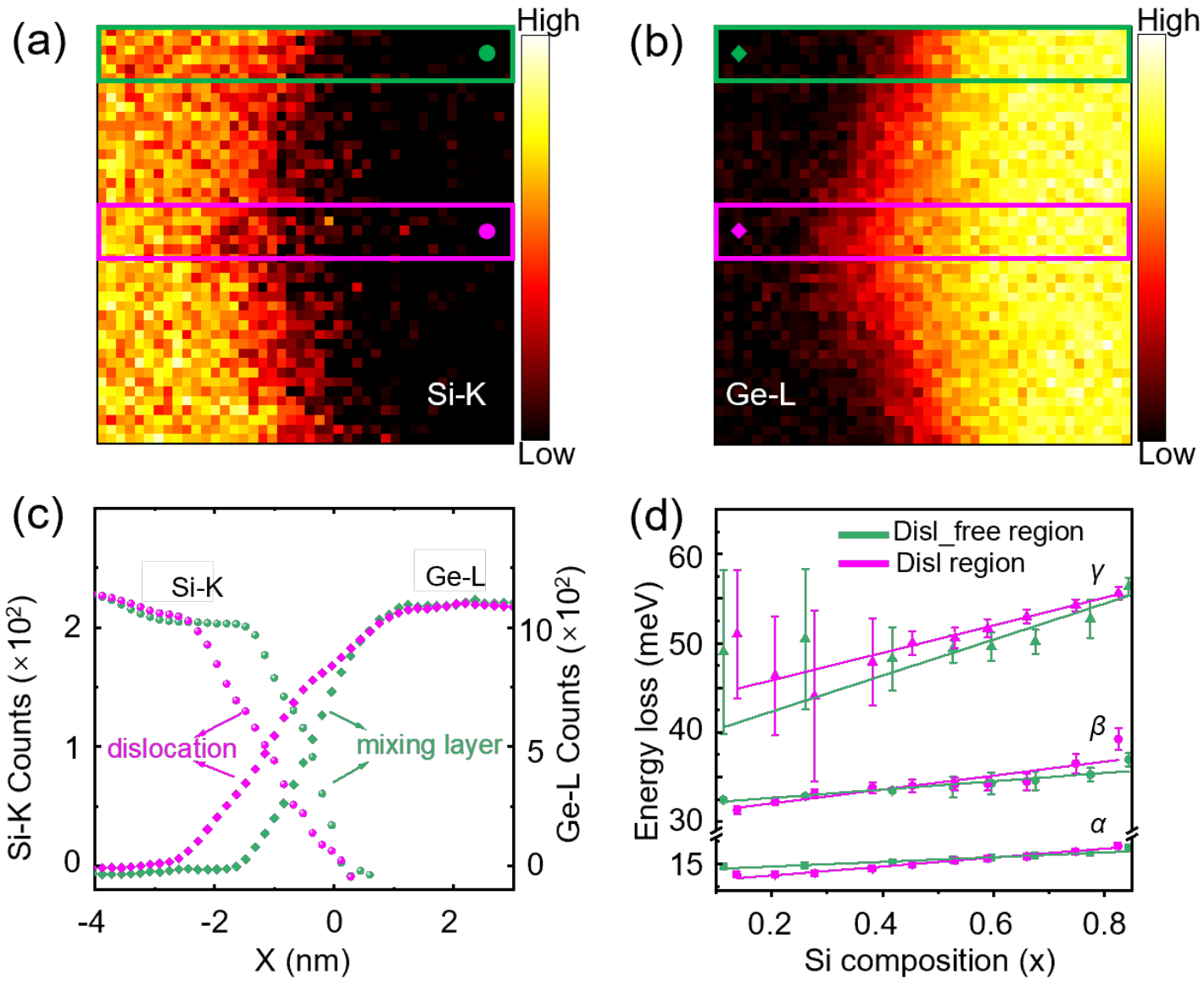

**Fig. S1. The element distribution at the interface**. **(a)** The intensity mapping of Si-*K* edge extracted from the core loss EELS. **(b)** The intensity mapping of Ge-*L* edge. **C** The intensity line profiles of the green region (disordered mixing layer with dislocation-free) and the magenta region (dislocation core) in **(a)** and **(b)**. **(d)** The phonon energy of $\alpha$, $\beta$ and $\gamma$ modes as a function of the Si composition ($x$), calculated from intensity ratio of Si-*K*/(Si-*K*+Ge-*L*), across the dislocation-free mixing layer (green) and the dislocation core (magenta). The error bar shows the standard deviations among neighboring 4×4 pixels.

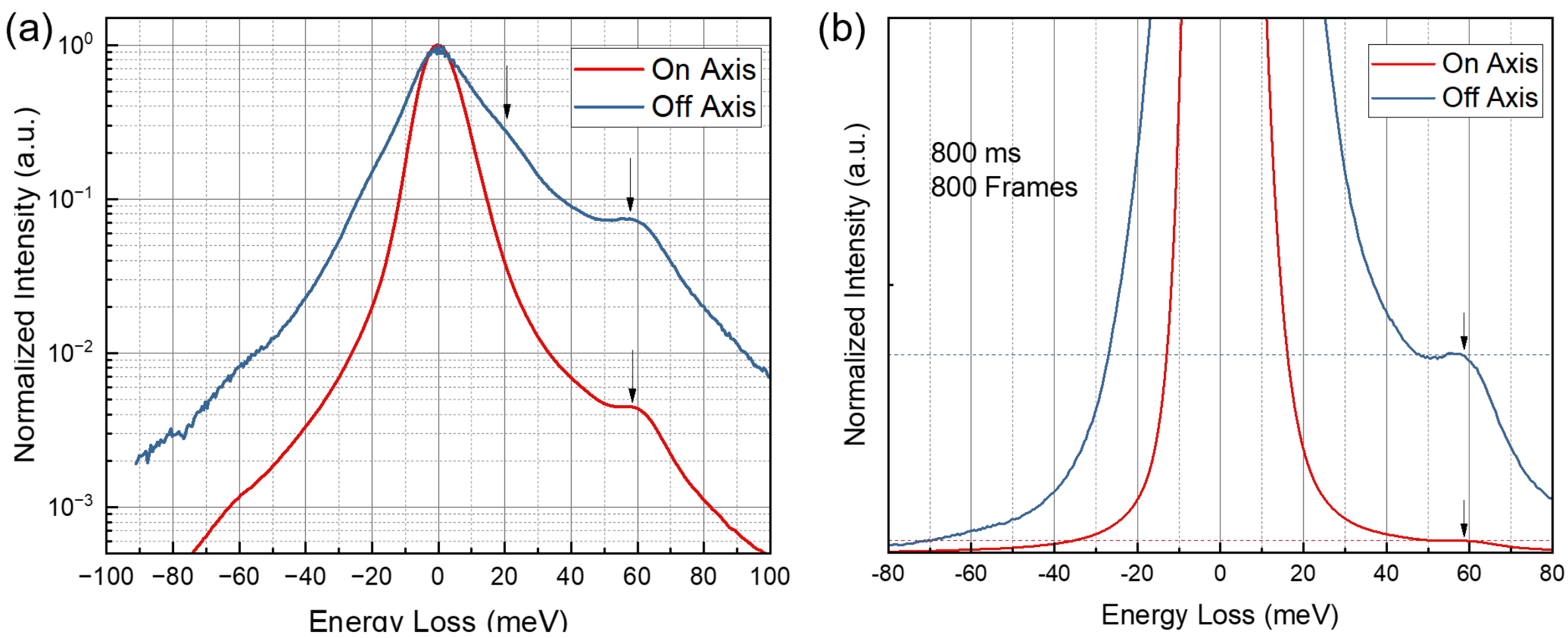

**Fig. S2. Comparison between on-axis and off-axis configurations. (a)** The spectra recorded under off-axis and on-axis configurations in log-scale. The spectra are normalized by the maximum intensity. **(b)** Partially enlarged image near baseline in linear scale. The off-axis spectra show stronger EELS intensity of phonon signals.

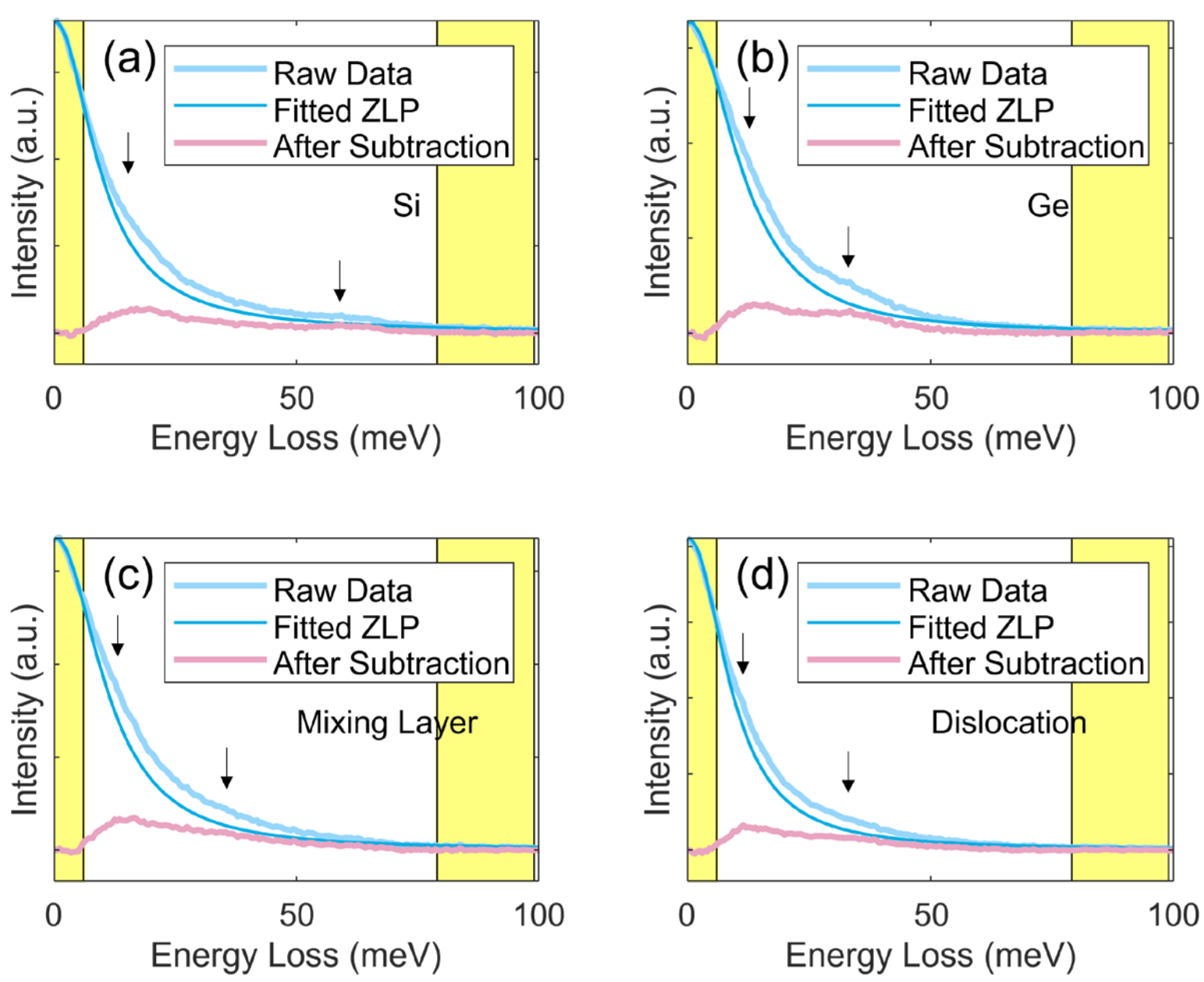

**Fig. S3. The background subtraction for experimental spectra. (a)** Bulk Si. **(b)** Bulk Ge. **(c)** The mixing layer at the interface without dislocation. **(d)** The dislocation at the interface. The thick light blue lines are the raw EEL spectra. Thin blue lines: the fitted ZLP using Pearson VII function. Pink lines: the signal after background subtraction. Strong phonon signals in low-frequency energy range can be seen even in raw spectra.

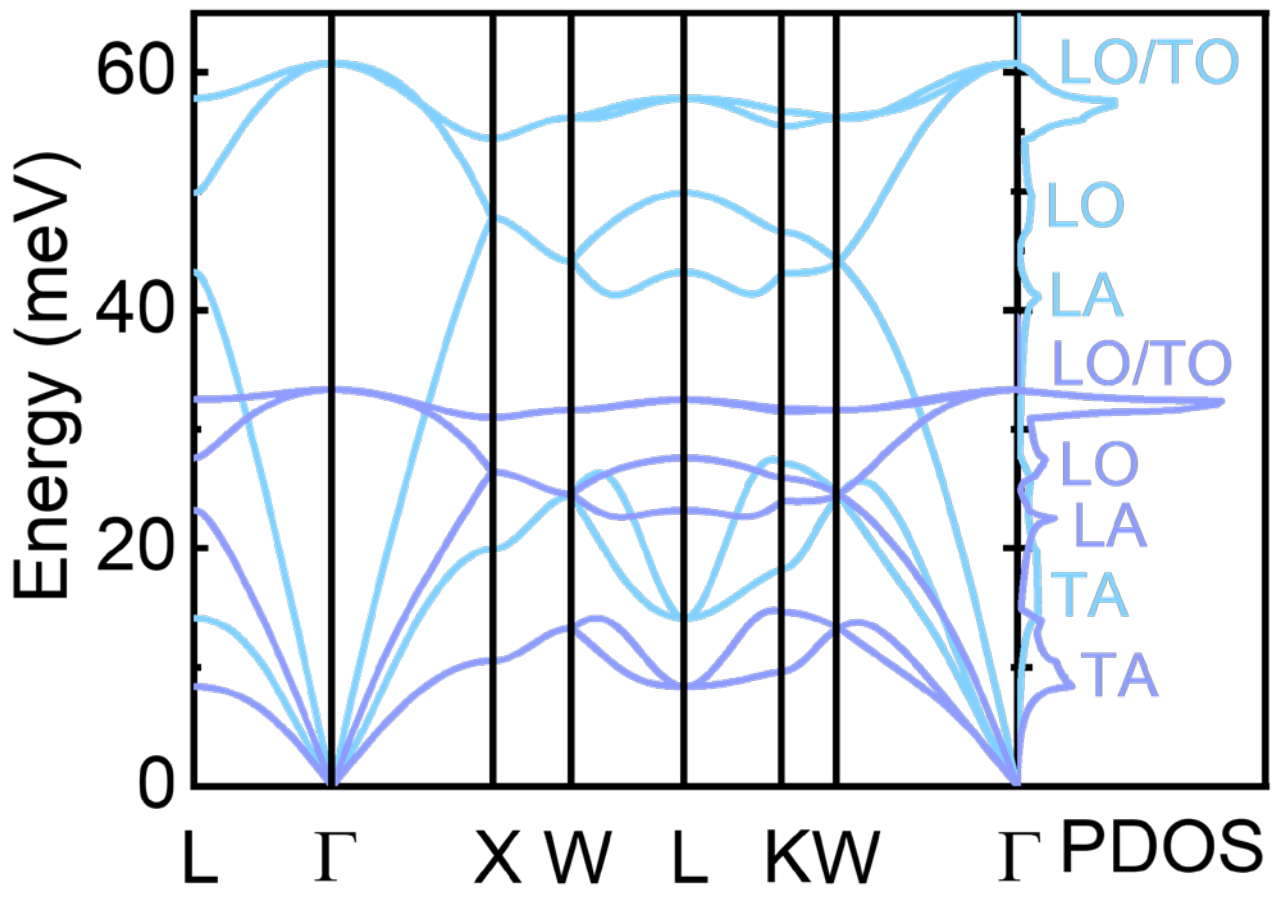

**Fig. S4.** Calculated phonon dispersions and corresponding PhDOS of Si (blue) and Ge (purple).

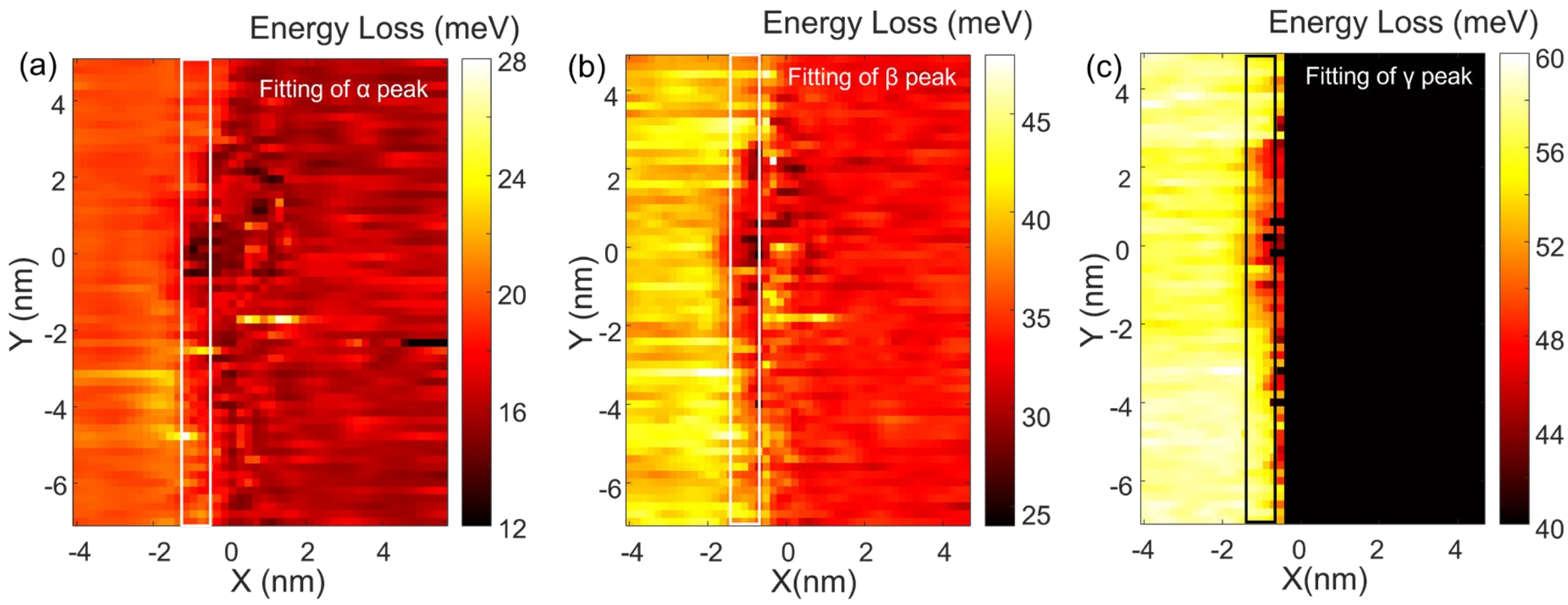

**Fig. S5.** The map of peak fitting for **(a)** $\alpha$, **(b)** $\beta$ and **(c)** $\gamma$ in full acquisition region of the spectrum image with a FOV of 12 nm × 9.6 nm. The rectangles denote the fitting area of data points shown in Fig. 2(f).

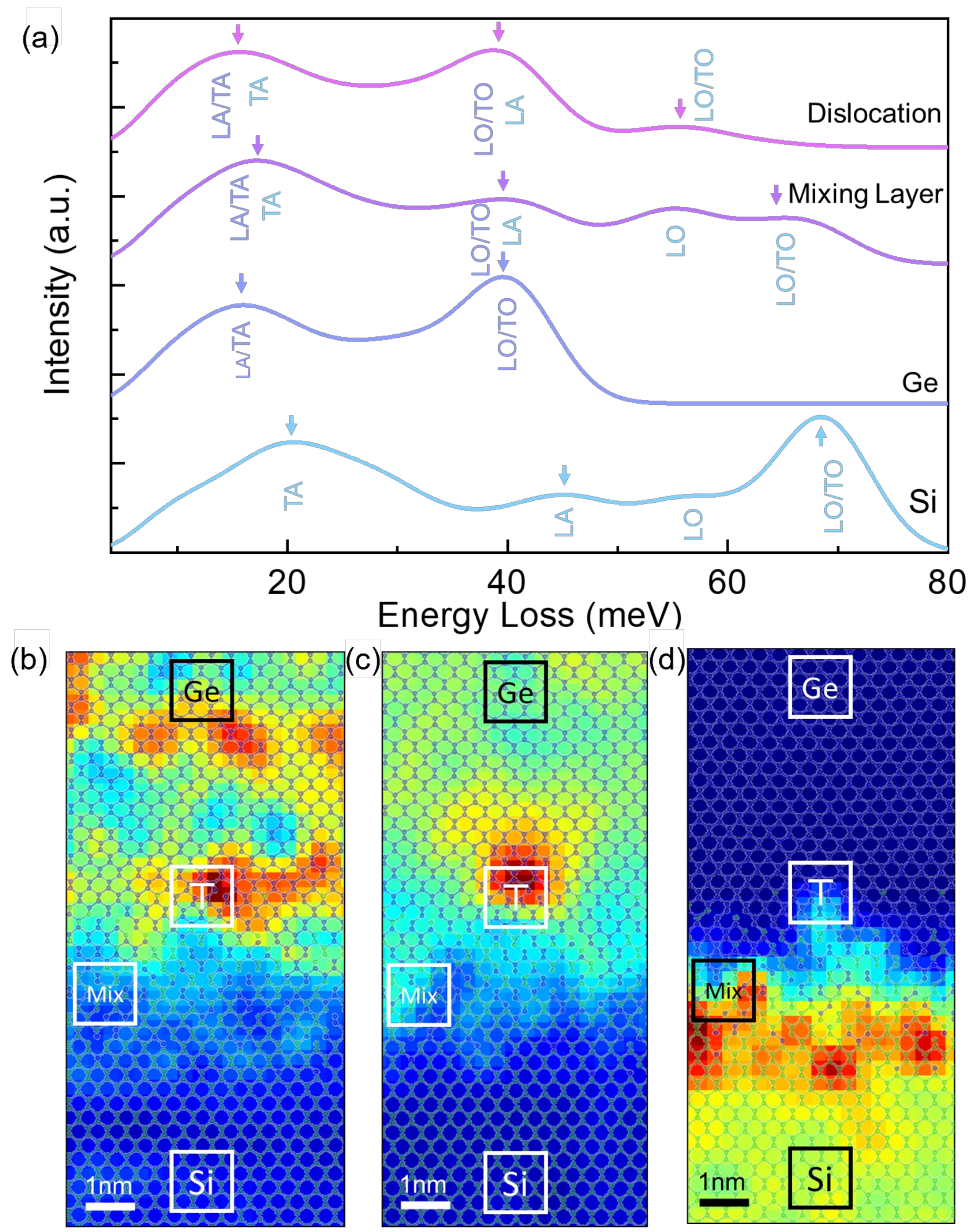

**Fig. S6. FRFPMS simulation of Si-Ge mixing interface with dislocation. (a)** Simulated spectra corresponding to rectangular area in **(b-d)**. **(b-d)** Energy filtered EELS spectrum image of 10-12 meV, 36-38 meV and 60-62 meV. The spectrum images b-d were processed with Gaussian blur of 3 × 3 pixels to eliminate the illusion caused by spatial undersampling. For b, the colormap was exponentiated to improve the contrast between the dislocation localized modes and Ge TA modes.

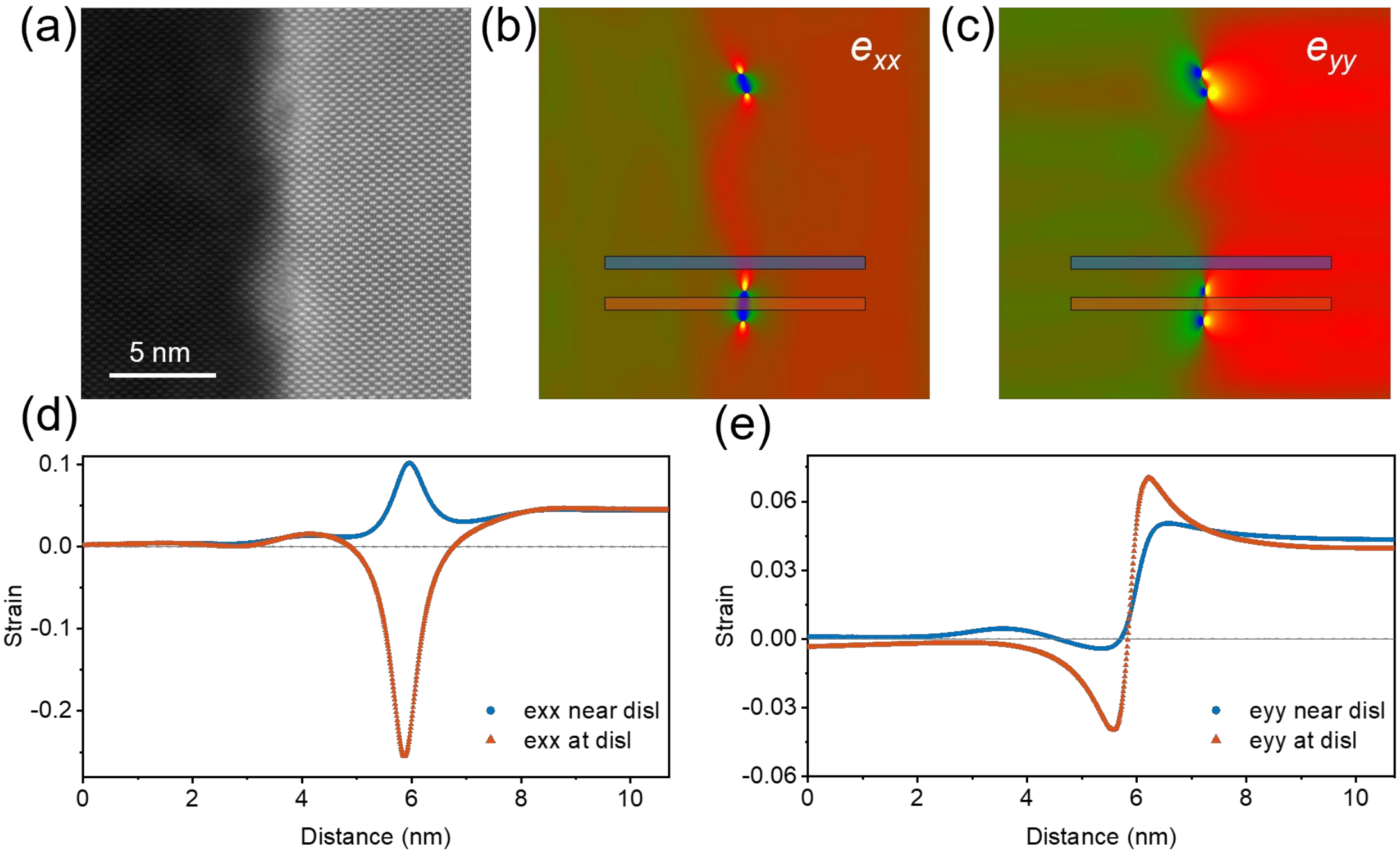

**Fig. S7. The strain distribution near Si/Ge dislocations. (a)** The HAADF image shows two adjacent dislocations at Si/Ge interface. **(b)** The horizontal strain $e_{xx}$ (perpendicular to the interface). **(c)** The vertical strain $e_{yy}$ (parallel to the interface). The strain concentrates at the dislocations. **(d-e)** Line profiles of **(d)** $e_{xx}$ and **(e)** $e_{yy}$ across the dislocation, highlighted by colored square in **(b)** and **(c)** (Red: strain at the dislocation; Blue: strain near the dislocation).

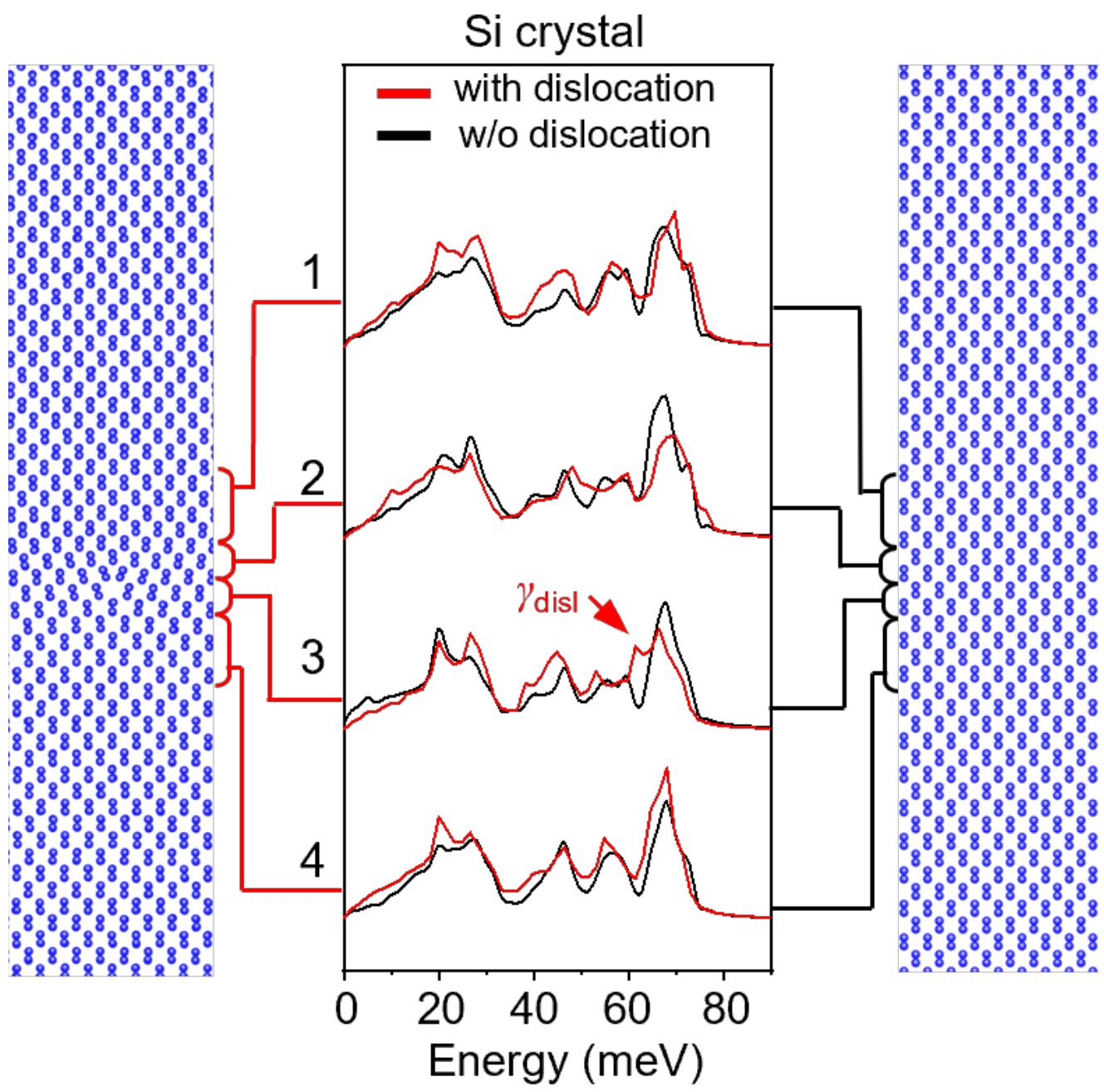

**Fig. S8. Molecular dynamics simulations of Si crystal with/without dislocation.** Schematic diagrams and calculated PPDOS of Si crystal. Each layer is the PPDOS of the atoms covered by the brace. The dislocation modes $\gamma_{disl}$ (the red arrow) experience a redshift compared to dislocation-free case (black).

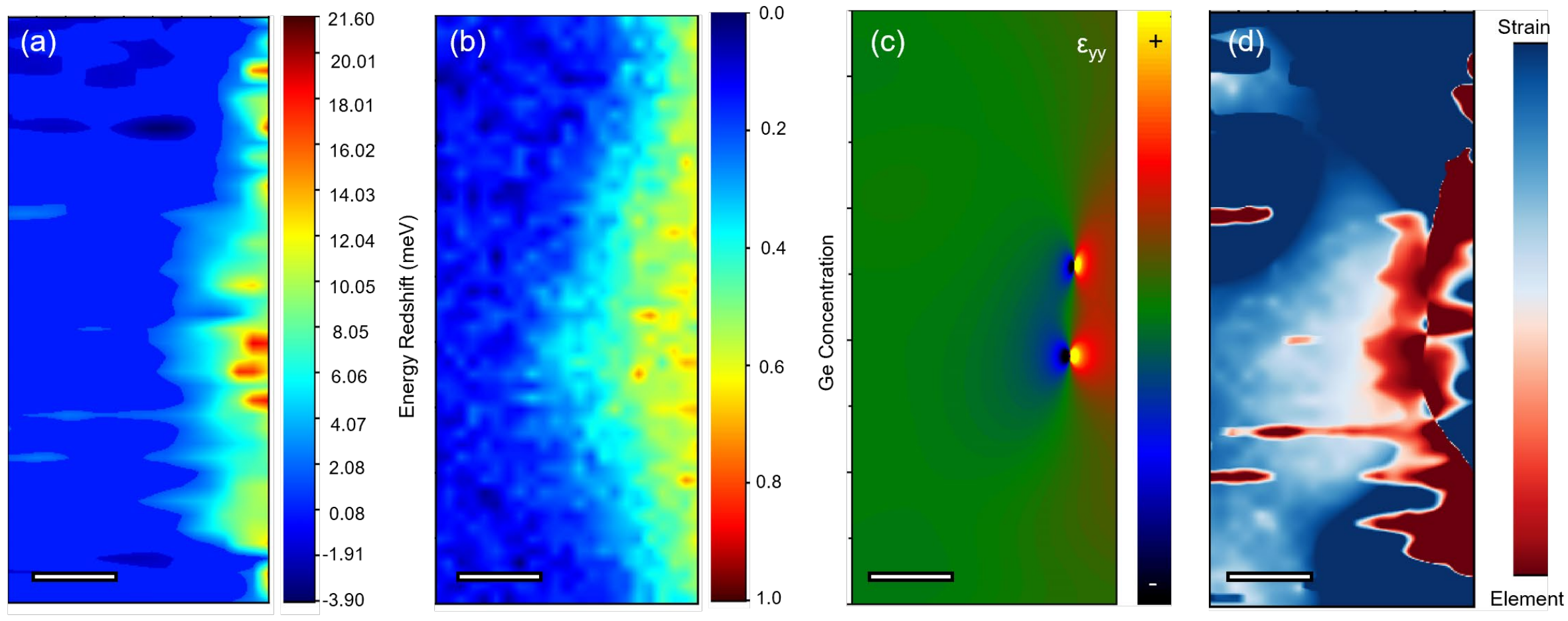

**Fig. S9. Decoupling the contribution of Ge concentration and strain to energy shift. (a)** The energy shift of $\gamma_{\mathrm{disl}}$ phonon mode with respect to the Si-LO bulk mode. **(b)** The Ge concentration from coreloss EELS of Si *K*-edge and Ge *L*-edge. **(c)** The $\varepsilon_{yy}$ strain in the vicinity of the dislocation based on atomically resolved HAADF image and GPA analysis. **(d)** A benchmark for comparing the contribution of Ge concentration and strain to energy shift. The benchmark is normalized to zero. If the benchmark is larger than zero, then strain contributes more to the energy shift and vice versa.

**Table S1. DOS difference across the interfaces/dislocations**

| Structure | DOS Area (Si side) | DOS Area (Ge side) | Area of overlap | Area of overlap (%) |
|---|---|---|---|---|
| Ge crystal without dislocation | 0.4166 | 0.4419 | 0.4117 | 98.84% |
| Ge crystal with dislocation | 0.4825 | 0.5101 | 0.4361 | 90.38% |
| Without dislocation without intermixing | 0.7223 | 0.5448 | 0.3727 | 51.59 % |
| With dislocation without intermixing | 0.5826 | 0.4779 | 0.3494 | 59.97 % |
| Without dislocation with Ge rich intermixing | 0.6119 | 0.5261 | 0.4218 | 68.94% |
| With dislocation with Ge rich intermixing | 0.6683 | 0.4919 | 0.3992 | 59.74% |

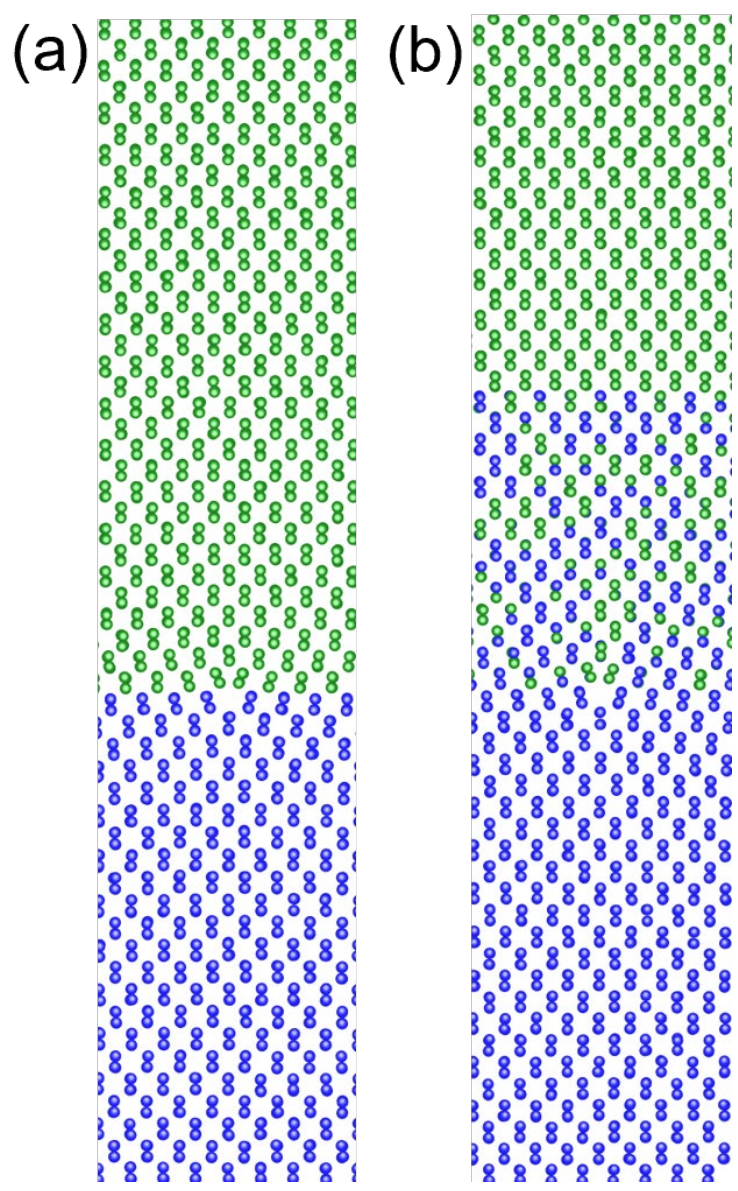

**Fig. S10. Dislocations facilitate increasing thermal conductivity at Si/Ge interface. (a)** The schematic diagrams of a Si/Ge interface without Si/Ge intermixing. **(b)** The schematic diagrams of a Si/Ge system with 50%Si-50%Ge intermixing.

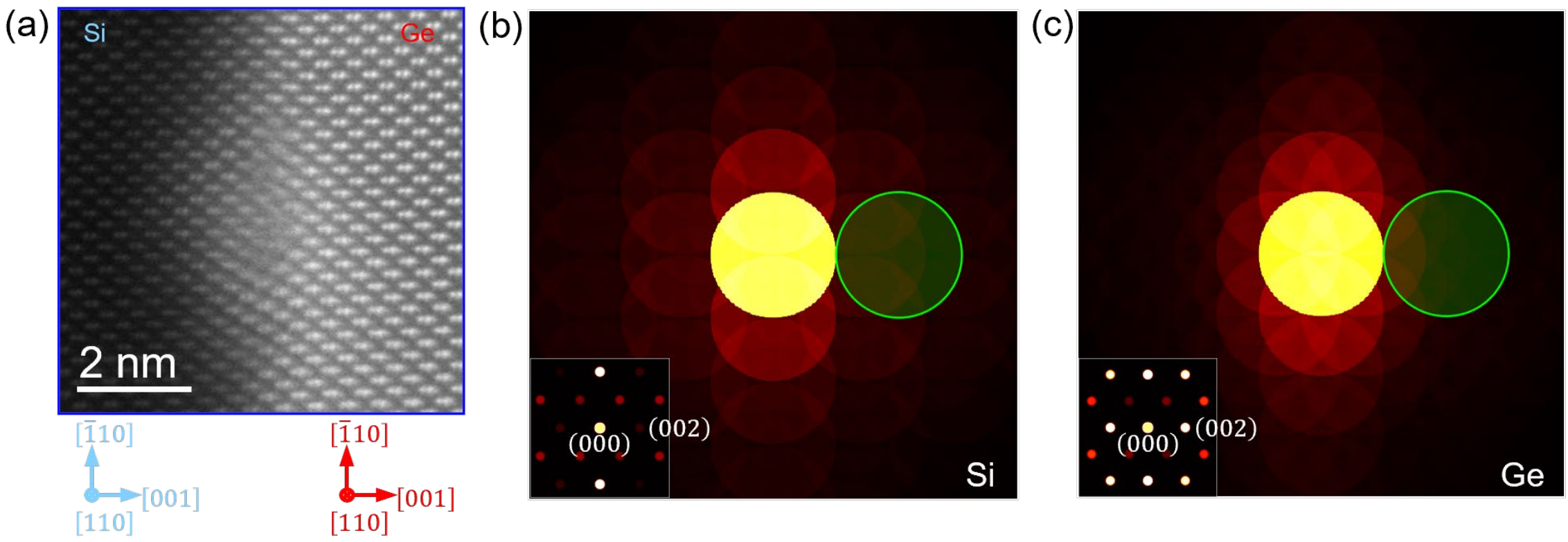

**Fig. S11. Atomic structure of interface (a) and the diffraction disks of Si (b) and Ge (c).** The red color indicates the superposition of diffraction disks along the [110] zone axis (convergence semi-angle: 25 mrad), while the green represents the EELS signal collection region (collection semi-angle: 25 mrad) included by the electron beam for off-axis experiments, which was moved off axis perpendicular to the interface with ~50 mrad.

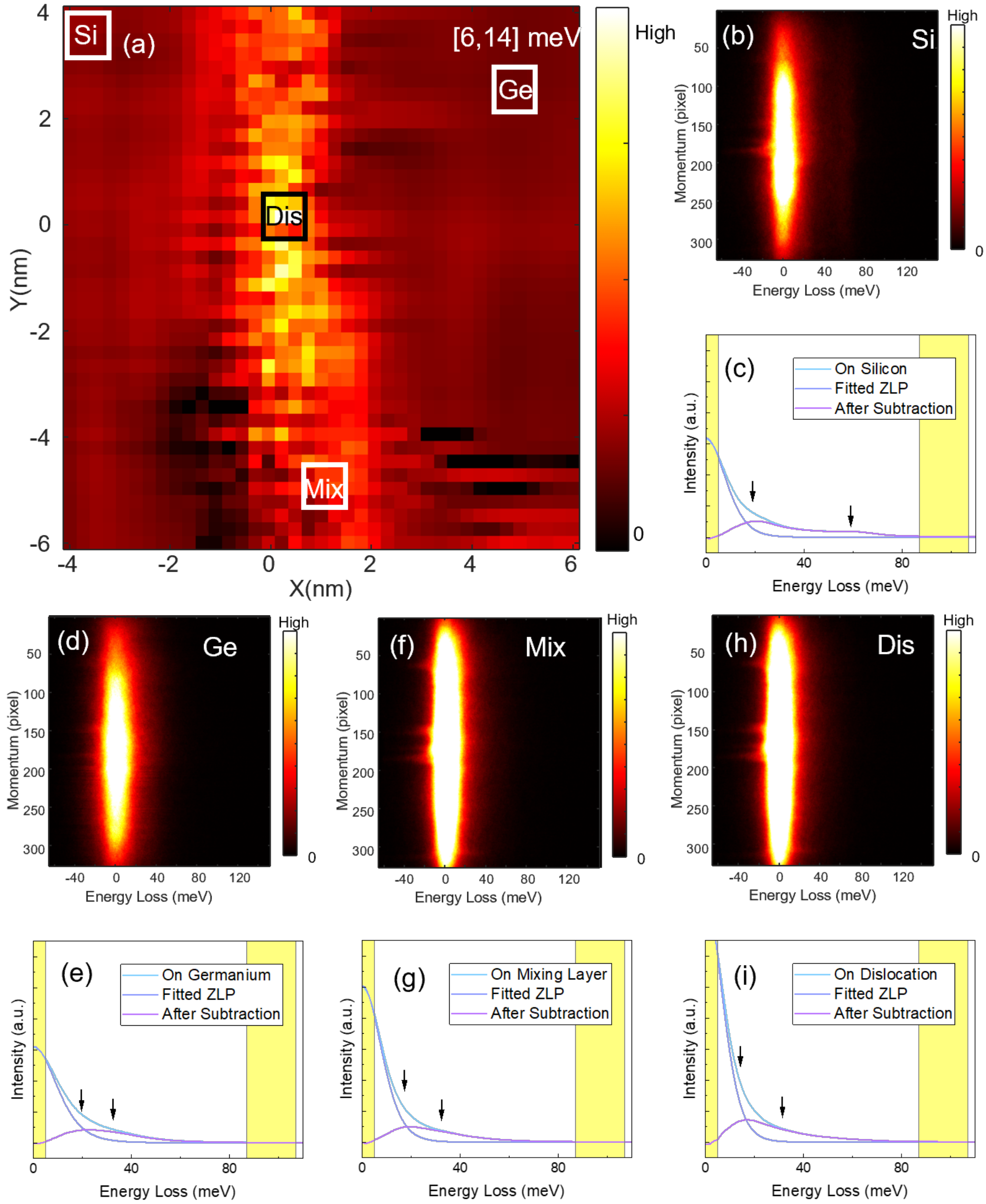

**Fig. S12. 2D-EELS image and EEL spectra. (a)** Spectrum image obtained by integrating signals of [6,14] meV. Four positions with 3 × 3 pixels are chosen to extract the 2D-EELS image and EEL spectra. **(b), (d), (f), (h)** 2D-EELS image from rectangular areas in **(a)**. Along the momentum direction, each profile is aligned (i.e. ZLP alignment) using maximum cross-correlation method. **(c), (e), (g), (i)** EEL spectra corresponding to **(b), (d), (f), (h)**.

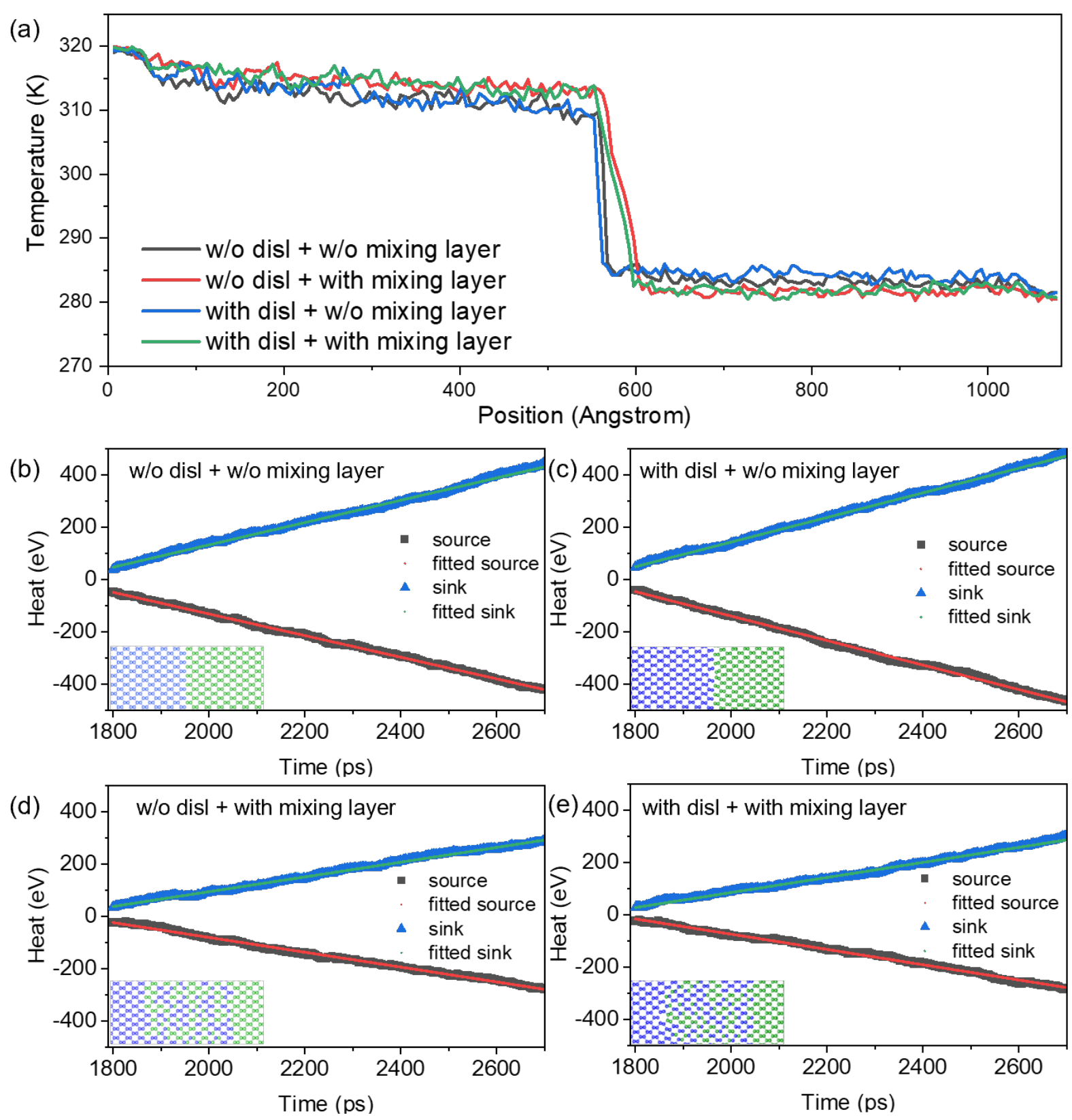

**Fig. S13. The temperature profile and heat current of NEMD simulations. (a)** Temperature profiles of NEMD simulations. **(b-e)** Accumulated heat of the heat source domain and heat sink domain varying as a function of the simulation time.

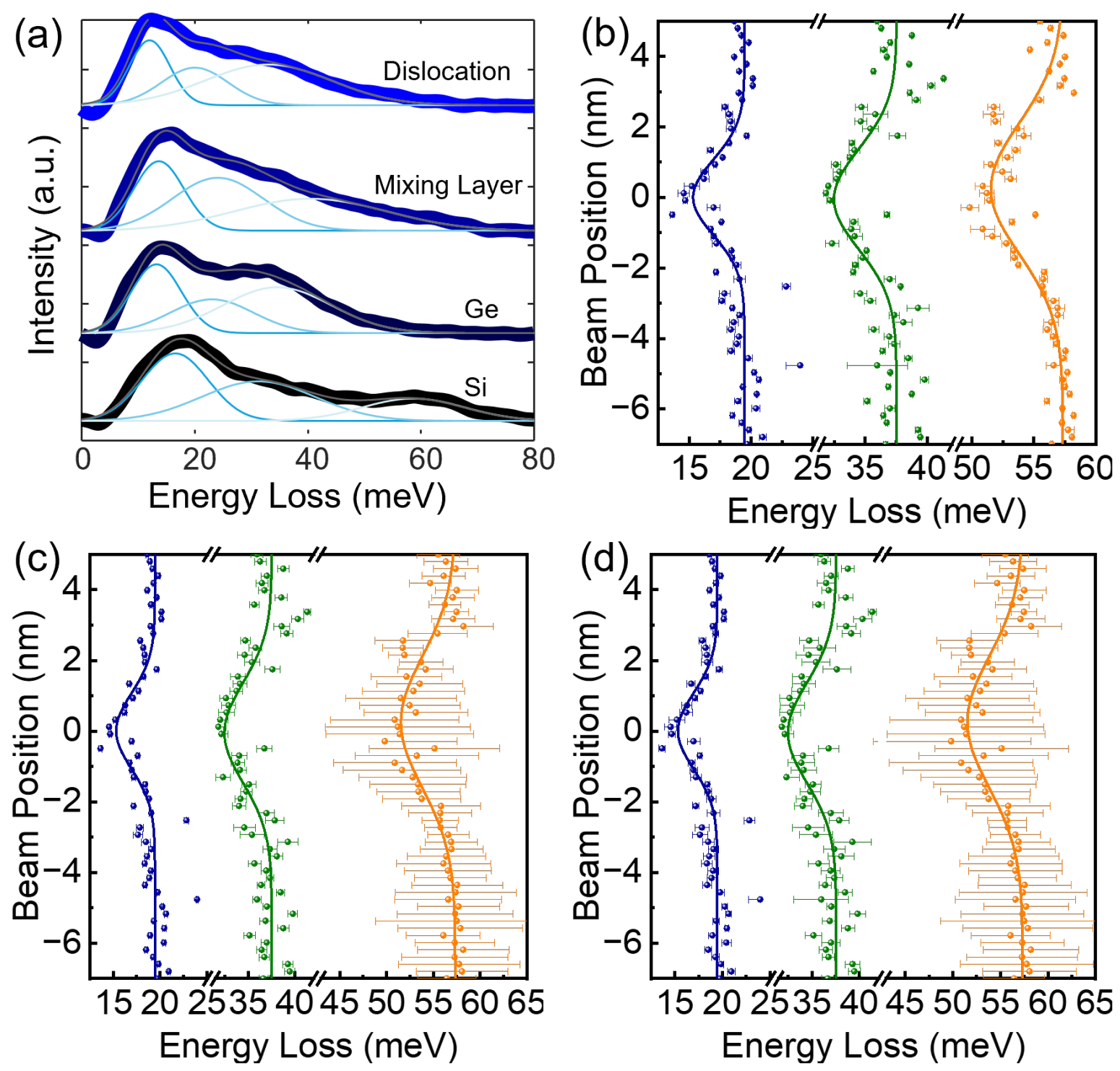

**Fig. S14. The multi-gaussian fitting results of EEL spectra. (a)** Multi-gaussian fitting of the EEL spectra acquired at different positions. The stand deviation error **(b)**, the Gaussian fitting error **(c)** and their combined uncertainty **(d)** for peaking fitting of phonon energies related to Figure 2.

# S2 Supporting Information: Relationship between element intermixing and phonon energy.

The element distribution maps extracted from core loss EELS (Si-*K* edge and Ge-*L* edge) are shown in **Fig. S1(a)** and **Fig. S1(b)**. The dislocation promotes the diffusion of Ge into the Si layer as the dislocation reduces activation energy for diffusion[1-3], leading to a larger diffusion depth. **Figure S1(c)** shows the line profile of Si composition and Ge composition across the mixing layer (green) and interface dislocation (magenta) along X direction denoted by rectangles in **Fig. S1(a)** and **Fig. S1(b)**, respectively. Approaching to the interface (at 0 nm) in the Si side, the Si composition becomes lower and the Ge composition becomes higher. Correspondingly, the phonon energies of $\alpha$, $\beta$ and $\gamma$ peaks gradually decrease.

**Figure S1(d)** shows the energy of $\alpha$, $\beta$ and $\gamma$ peaks as functions of the Si composition ($x$), intensity ratio Si-*K*/(Si-*K*+Ge-*L*), in the mixing layer. The energies of $\alpha$, $\beta$ and $\gamma$ peaks are approximately linearly related to the Si composition, with slopes of 2.61±0.23 meV, 4.66±0.62 meV and 20.16±4.04 meV in the dislocation-free (green) region; 5.02±0.24 meV, 7.79±1.03 meV and 15.41±1.05 meV in the dislocation (magenta) region, respectively. There are two major competitive effects in the mixing layer that can change the slopes. One is that as Si composition increases, the bond length decreases, which leads to an upshift of vibrational energy. The other is that the disorder-induced shorter order range leads to vibrational energy downshifts with respect to an ideal lattice[4]. That is to say, as Si composition increases, the disorder-induced upshift of vibrational frequency of Si-Si bond and the downshift of vibrational frequency of Ge-Ge bond both exist. The γ mode includes the contributions of Si-Si bonds and Si-Ge bonds but not Ge-Ge bonds, so its slope is the largest. The α mode is below 18 meV, containing mainly the contribution of Ge-Ge bonds, therefore its slope is minimal. For slopes in the mixing layer without dislocation, they were consistent with that reported in literatures[5,6]. However, in the dislocation region, the $\gamma$ peak has a significantly smaller slope and larger energy loss than those in dislocation-free region, indicating the dislocation has more contribution on the $\gamma$ mode and less effect on phonons with lower energy ($\alpha$ and $\beta$ mode), which imply the contribution of defect on the phonon spectra.

At the interface, the α peak connects the Si-TA mode and Ge-TA mode, the $\beta$ peak connects the Si-LAO mode and Ge-AO mode. Therefore, the α peak and $\beta$ peak gradually convert Si phonons to Ge phonons through the mixing layer, which facilitates the phonon transport through the interface and is beneficial to increase ITC[7]. On the other hand, the disorder-induced phonon backscattering reduces phonon transmittance and reduce ITC was also reported[7,8]. Whether the defect enhances or reduces ITC depends the competition of these effects[7].

# S3 Supporting Information: Advantages of the off-axis configuration compared to the on-axis configuration.

Note that the previous works[9,10] usually used on-axis configurations to record Si vibrational spectra, intending to achieve higher intensity. In our case, the off-axis configuration is chosen because it can offer higher signal-to-ZLP ratio of phonon signals especially for the low frequency phonons, as demonstrated in **Fig. S2**.

# S4 Supporting Information: Methods for decoupling the contribution of Ge segregation and strain to the energy shift.

**1. Data registration**

The special maps of the energy-shift image and the Ge concentration were firstly interpolated into the same resolution of the GPA image using bilinear method. Then, using the energy-shift image as a template, the mutual information (MI) between the template and the two data map were calculated[11]. MI is defined using information entropy of the data map. The information entropy of a data map is defined using

$$H(Y) = -\sum_{i=0}^{N=1} p_i log p_i \quad (1)$$

where $p_i$ can be calculated using the distribution function of the data. Then the combination entropy between two data maps can be defined using similar method, while *p(x,y)* can be calculated using joint distribution function of two data maps:

$$H(X,Y) = -\sum_{x,y} p_{xy}(x,y) log p_{XY}(x,y) \qquad (2)$$

Then MI can be calculated using

$$MI(X,Y) = H(X) + H(Y) - H(X,Y) \qquad (3)$$

We slide the template on the reference data map and calculate MI of different pixels. The data maps achieve registration when MI get the largest number.

**2. Weight coefficient for decoupling the effect of strain and composition**

A bilinear model was firstly constructed:

$$z(x,y) = \alpha(bx) + (1-\alpha)(ky) \qquad (4)$$

where $z$ is the net energy shift, the first term on right-side $\alpha(bx)$ elucidates energy shift caused by strain while the term $(1-\alpha)(ky)$ means energy shift caused by element interdiffusion. For the first term, the factor $b$ can be calculated using the model described in previous study[12]. For the second term, the factor $k$ can be fitted linearly using data in **Fig. S1**. $\alpha$ can be seen as a weight coefficient. When $\alpha < 0.5$, the elemental interdiffusion dominate the energy shift and vice versa. An square area of 5 ×5 pixels was chased as a dataset to fit the model in Eq 4. The $\alpha$ of each place on the energy map can be collected by sliding the square all around the data map.

# S5 Supporting Information: Results of decoupling the contribution of Ge segregation and strain to the energy shift.

In the Si side (compressive stress region) at the dislocation core, the energy shift of $\gamma_{disl}$ phonon mode is dominated by elemental component segregation. Even inside the dislocation core, elemental segregation also has a great contribution to the red shift of $\gamma_{disl}$ phonon. In the region slightly away from the dislocation core, there is a competition (white region in **Fig. S9(d)**) between these two components.

It also should be noted that the linear model used in the decoupling method is simple and may not accurately describe the complex real coupling of strain and elemental distribution. The fitting of element proportion and the phonon energy also have evitable uncertainty. Even though, it provides a qualitative and intuitive way to understand the coupling effect between strain and elemental segregation.

## S6 Supporting Information: Verifying low frequency signals using 2D EELS images.

**Figure S12** shows the 2D-EELS images and corresponding EEL spectra. Although there is some tiny spiky spurious signal below 20 meV, they contribute little to the total signal after alignment and summing, which is also evident by the smooth and continuous curve of the signal in the low frequency region. The main feature, i.e., the enhanced intensity with energy below 20 meV (redshift of the α-phonon peak) at the dislocation core, is consistent with **Fig. 2(a)** and **Fig. 1(d)**.

# S7 Supporting Information: References